\documentclass[aps, twocolumn,
superscriptaddress,showpacs,showkeys,amsmath,amssymb,floatfix]{revtex4}
\usepackage{color}
\usepackage{graphicx}
\usepackage{epsfig}
\usepackage{dcolumn}
\usepackage{bm}
\usepackage{mathtools}
\usepackage{amssymb}
\usepackage{bmpsize}
\usepackage{calrsfs}
\DeclareMathAlphabet{\pazocal}{OMS}{zplm}{m}{n} 
\usepackage{amsmath}
\usepackage{subfigure}
\newcommand{\be}{\begin{equation}}
\newcommand{\ee}{\end{equation}}
\newcommand{\bea}{\begin{eqnarray}}
\newcommand{\eea}{\end{eqnarray}}

\newcommand{\hg}{\hat g}

\newcommand{\tc}{{\tilde c}}
\newcommand{\tg}{{\tilde g}}

\newcommand{\tw}{{\tilde w}}

\newcommand{\pro}{\partial}

\newcommand{\bfA}{{\mathbf A}}

\newcommand{\bfr}{{\mathbf r}}

\newcommand{\bfw}{{\mathbf w}}

\newcommand{\ba}{\begin{array}}
\newcommand{\ea}{\end{array}}

\newcommand{\nn}{\nonumber}

\newcommand{\Ab}{\pazocal{A}}

\begin{document} 
\title{Color structure of quantum $SU(N)$ Yang-Mills theory}
\author{D.G. Pak}
\affiliation{Institute of Theoretical Physics, Chinese Academy of Sciences, Beijing 100190, China}
\affiliation{School of Education, Bukkyo University, Kyoto 603-8301, Japan}
\affiliation{Chern Institute of Mathematics, Nankai University, Tianjin 300071, China} 
\author{Takuya Tsukioka}
\affiliation{School of Education, Bukkyo University, Kyoto 603-8301, Japan}
\begin{abstract}
Color confinement  is the most puzzling phenomenon in the theory of strong interaction, quantum chromodynamics,
  based on  a quantum $SU(3)$ Yang-Mills theory. 
The origin of color confinement supposed to be intimately related to non-perturbative features of the non-Abelian gauge 
theory, and touches very foundations of the theory. We revise basic concepts underlying QCD concentrating mainly 
on concepts of gluons and quarks and color structure of  quantum states. Our main idea 
is that a Weyl symmetry is the only color symmetry which determines all color attributes of quantum states and 
physical observables. We construct an ansatz for classical  Weyl symmetric dynamical solutions in $SU(3)$
Yang-Mills theory which describe one particle color singlet quantum states for gluons and quarks.
Abelian Weyl symmetric solutions provide microscopic structure of a color invariant vacuum and vacuum gluon condensates. 
This resolves a problem of existence of a gauge invariant and stable vacuum in QCD. 
Generalization of our consideration to $SU(N)$ $(N=4,5)$ Yang-Mills theory
implies that the color confinement phase is possible only in $SU(3)$ Yang-Mills theory. 
 \end{abstract}
\maketitle
     
A gauge principle is a fundamental guiding principle which provides construction of
strict mathematical theories of fundamental interactions \cite{Weyl1929}. A quantum chromodynamics (QCD)
is a well established theory of strong interaction which represents a quantum theory of classical 
Yang-Mills gauge theory  with a color group $SU(3)$ \cite{Yang-Mills}, the strict formulation of which is unknown so far \cite{Mill2000}.
The gauge symmetry plays an important role in determination of dynamical laws in the theory.
Formally,  the gauge symmetry looks redundant, since it leads to unphysical pure gauge degrees of freedom
which are usually  removed from the theory by applying a qauge fixing procedure 
in such  a way that final physical quantities do not depend on a chosen gauge.
So that, the local gauge symmetry governs the time evolution of the physical system, but can not serve as a symmetry 
describing color attributes of fundamental particles and quantum states. On the other hand, a global color 
symmetry $SU(3)$ can not be a real symmetry of particles as well, since (i) color particles are not observed due to color 
confinement phenomenon, (ii) there is no color charge conservation law like the electric charge conservation, 
and (iii) an escape from such paradoxical situation like that
``a color symmetry exists but it is spontaneously broken'' fails because it is in contradiction with  
the gauge invariant  structure of the QCD vacuum which provides the color confinement \cite{polyakov77,thooft81}. 

 In the present paper we demonstrate that a Weyl group of the structure group $SU(N)$ 
represents a finite color symmetry  which is the only color symmetry available after gauge fixing and which
describes color properties of quantum states and physical observables in a consistent manner with the color confinement phenomenon.

In Section I we consider a general structure of quantum Yang-Mills theory and main requirements to 
classical and quantum fields necessary for construction of one particle color singlet quantum states. 
In Section II we consider a minimal ansatz for stationary axially symmetric solutions of magnetic type in $SU(2)$ Yang-Mills theory.
We show that solutions admit a global color symmetry $SO(2)$ which remains after the gauge fixing and
 leads to a degenerated vacuum structure. 
This implies that $SU(2)$ Yang-Mills theory does not admit the color confinement phase.
An $SU(3)$ Yang-Mills theory with fermions, i.e., quantum chromodynamics with one flavor quark, is considered 
in Section III. We verify uniqueness of the Weyl symmetric ansatz and show that there are two possible classes
 of Weyl symmetric solutions, but only one class is consistent with the Weyl symmetric structure for quarks.
Our approach is non-perturbative in a sense that the Weyl symmetric structure of solutions is consistent
with the whole non-Abelian structure of equations of motion, and our ansatz describes exact solutions to equations of motion. 
Applying a non-trivial Abelian projection defined by the Weyl symmetric ansatz we obtain unexpected results: 
all Weyl symmetric solutions for gluons and quarks belong to irreducible Weyl representations leading to
color singlet one particle states contrary to commonly accepted view that gluons and quarks represent color particles 
as members of color octet and triplet respectively.
In Section IV we generalize our results to a case of $SU(4)$ Yang-Mills theory and construct the most general Weyl symmetric ansatz.
The obtained ansatz defines a two parametric family of solutions, so that vacuum solutions are degenerated. In general, this leads 
 to spontaneous color symmetry breaking.
Generalization to $SU(5)$ Yang-Mills theory and consclusions are contained in the last section.
      
\section{Structure of quantum Yang-Mills theory}

We consider a standard Lagrangian of $SU(N)$ ($N=2,3,4,5$) Yang-Mills theory in the presence of one 
complex multiplet of Dirac fermions  in the fundamental representation of $SU(N)$
\bea
&&{\cal L}={\cal L}_{YM}+{\cal L}_f, \nn \\
&&{\cal L}_{YM}=-\dfrac{1}{4} (F_\mu^a)^2, \nn \\
&&{\cal L}_{\rm f}= \bar \Psi \Big[ i \gamma^\mu (\pro_\mu -\dfrac{i g}{2} A_\mu^a \lambda^a)-m\Big]\Psi.
\eea
where $(\mu,\nu=0,1,2,3; a=1,2,...,N^2-1)$. Corresponding equations of motion describe a coupled 
system of a gluon field $A_\mu^a$ and Dirac spinor fields $\Psi_\alpha$ ($\alpha=1,2...N$)
\bea
&&(D^\mu \vec F_{\mu\nu})^a= -\dfrac{g}{2} j^a \equiv -\dfrac{g}{2}  \bar \Psi \gamma_\nu \lambda^a\Psi, \label{eqA}\\
&& \Big[ i \gamma^\mu (\pro_\mu -\dfrac{i g}{2} A_\mu^a \lambda^a)-m\Big]\Psi=0. \label{eqPsi}
\eea
A general quantization formalism based on functional integral provides 
a principal structure of quantum $SU(N)$ Yang-Mills theory.
Following this approach a gauge connection (gauge potential) $A_\mu^a$
is split into two parts
\bea 
&& A_\mu^a=A_\mu^{cl}+A_\mu^{quant},
\eea
where the quantum field $A_\mu^{quant}$ describes quantum fluctuations and stands for  arbitrary field configurations 
over which the functional integraton is performed. The integration over quantum fields represents summation over all gauge 
equivalence classes of the gauge fields which are extracted by a gauge fixing procedure in a gauge invariant manner.
So that for description of quantum fields one can use an arbitrary complete basis in the space of vector and spinor fields 
for gluons and quarks respectively. 
In a particular, in perturbative quantum chromodynamics (QCD) a basis of free plane waves is usually applied
which leads to concepts of eight color gluons and three color quarks. Certainly, such defined gluons and quarks
provide proper description of quantum gluons and quarks in internal lines and loops of Feynman diagrams and 
represent virtual quantum objects which are not observable directly. 
Observable particles are represented by a complete set of
classical dynamical solutions which produce a Hilbert space of quantum states.
A spectrum and properties of one particle quantum states depend on a chosen gauge fixing procedure 
and selected solution basis. Due to lack of a linear superposition principle in the non-Abelian theory a choice of
different solution bases lead to non-equivalent particle contents or even unphysical particles. In a particular, the basis of plane wave 
solutions leads to unphysical concepts for gluon and quark particles. Therefore, first of all one should define necessary
requirements to classical solutions which provide consistent definitions of corresponding particles or quantum states.

In QCD the classical field $A_\mu^{cl}$  plays another important role, due to the presence  of a nontrivial vacuum \cite{savv}.
Such a vacuum generates non-zero vacuum gluon and quark condensates, and we need some classical solutions which describe
microscopic structure of the vacuum condensates. This implies a well known long-standing problem of construction of a vacuum fields which are 
stable against quantum fluctuations \cite{N-O}. Moreover, it has been observed by Polyakov, \cite{polyakov77}, that color confinement phenomenon
is provided by a color invariant vacuum. `t Hooft conjectured that such vacuum structure is described by Abelian fields
according to a so-called Abelian projection of QCD, which was confirmed later by numerical study \cite{abeldom1, abeldom3}. 
Such a resolution to the vacuum problem is not merely satisfactory, because
obviously any non-trivial vacuum solution is not invariant under color transformation. 
A possible way to resolving this problem is given by H. Weyl emphasizing in his book \cite{Weyl1952} the importance of symmetric 
configurations invariant under a subgroup of the initial symmetry group. Following this idea, the problem of construction of an
invariant vacuum solution reduces to finding irreducible representations of a finite color subgroup of $SU(3)$ which admit
a one-dimensional space of field variables. Then such one-dimensional irreducible representations will define color singlet vacuum
and quantum states. 

We impose the following main requirements to classical solutions describing one particle quantum states for gluons and quarks:\\
(a) the consistency with quantum mechanical principles implies that structure elements, elementary particles, 
must be vibrating at microscopic space-time level \cite{niel-oles,amb-oles2}. 
This leads to a condition that classical solutions describing fundamental particles and non-trivial vacuum must be time dependent. 
By this way one can obtain microscopic description of vacuum quark and gluon condensates;\\
(b) solutions must be exact solutions to non-linear equations of motion. Solutions 
obtained in some approximation (like the plane waves in perturbation theory) may not exist as strict solutions in the full theory
and cause inconsistency with non-perturbative features;\\
(c) solutions must possess clasiical and quantum stability. Some unstable classical solutions are admissible as physical ones
since they may correspond to metastable states. However, the quantum stability against quantum fluctuations is obligatory
since it guarantees stability of the vacuum in the theory;\\
(d) solutions must admit localization of particles, in particular the energy, total angular momentum must be conserved in a 
finite space regions, since all known hadrons are localized objects;\\
Strictly speaking, in quantum theory one should consider classical solutions to equations of motion 
obtained from the functional of quantum effective action $\Gamma_{eff}$ which contains quantum corrections. 
In the present paper we constrain our consideration 
within the quasiclassical consideration neglecting quantum corrections. An important role of quantum corrections in providing 
localized solutions is demonstrated in \cite{PRD1}.\\
(e) any classical solution breaks the initial color symmetry. To classify color attributes quantized solutions
one needs solutions which still possess a symmetry subgroup of $SU(N)$.
Our main assumption is that a Weyl symmetry provided by the Weyl group of $SU(N)$ 
determines color properties of quantum states and color confinement structure. 

\section{Color structure of $SU(2)$ Yang-Mills theory}

A generalized 
axially symmetric Dashen-Hasslacher-Neveu (DHN)  ansatz for stationary solutions of magnetic type is defined by the following 
non-vanishing components of the gauge potential \cite{plb2018} 
\begin{align}
A_t^2&=K_0(r,\theta,t),& A_r^2&= K_1(r,\theta,t),~~A_\theta^2= K_2(r,\theta,t), \nn \\
A_\varphi^3&=K_3(r,\theta,t), & A_\varphi^1&= K_4(r,\theta,t), \label{genDHN}
\end{align}
 where non-vanishing component fields of the gauge potential are presented by one Abelian magnetic potential, $K_3$,
 and by four off-diagonal fields $K_{0,1,2,4}$ in standard spherical coordinates ($r, \theta, \varphi$).
 One can verify, the ansatz admits a non-trivial residual local $U(1)$ symmetry with a gauge parameter $\lambda(r,\theta,t)$
\begin{align}
K'_0&=K_0+\pro_t \lambda,~~K'_1=K_1+\pro_r \lambda,~~K'_2=K_2+\pro_\theta \lambda, \nn \\
K_3'&=K_3 \cos  \lambda+K_4\sin  \lambda, \nn\\
K_4'&=K_4 \cos  \lambda-K_3 \sin  \lambda.\label{residual} 
\end{align}
Due to the presence of $U(1)$ local symmetry one has pure gauge degrees of freedom which can be 
eliminated from the space of classical solutions by fixing a gauge. It is suitable to add additional  Lorenz type gauge fixing terms 
${\cal L}_{g.f} $ to the original Yang-Mills Lagrangian ${\cal L}_{YM}$ 
 \bea
{\cal L}_{\rm gf}=-\dfrac{1}{2} (\pro_t K_0-\pro_r K_1-\dfrac{1}{r^2} \pro_\theta K_2)^2. \label{gfterm}
\eea
With this the ansatz (\ref{genDHN}) is consistent with all equations of motion corresponding to the original Yang-Mills Lagrangian 
with gauge fixing terms and leads to five independent equations for the fields $K_{0,1,2,3,4}$.
One has still a global $SO(2)$ symmetry with a constant parameter $\lambda$
which can be fixed by imposing a constraint on azumuthal components of the gauge potential 
\bea
K_3=\tc_3 K_4,\label{const}
\eea
where $\tc_3$ is an arbitrary real constant.
After imposing constraint (\ref{const}) the five equations for fields $K_{01,2,3,4}$ reduce to
 four independent second order hyperbolic 
differential equations for fields $K_{0,1,2,3}$
\bea
&&r^2 \pro_t^2 K_1-r^2 \pro_r^2 K_1-\pro_\theta^2 K_1+2r(\pro_t K_0-\pro_r K_1) \nn \\
&&~~~+\cot\theta(\pro_r K_2-\pro_\theta K_1)+\dfrac{9}{2}\csc^2\theta K_4^2 K_1=0,\label{eq1} \\
&&r^2\pro_t^2 K_2-r^2 \pro_r^2 K_2-\pro_\theta^2 K_2+r^2 \cot\theta(\pro_t K_0 \nn \\
&&~~~-\pro_r K_1)-cot\theta \pro_\theta K_2+\dfrac{9}{2}\csc^2\theta K_4^2 K_2=0,  \label{eq2} \\
&&r^2 \pro_t^2 K_4-r^2 \pro_r^2 K_4-\pro_\theta^2 K_4+\cot\theta\pro_\theta K_4\nn \\
&&~~~~~~~~~~~~+3r^2(K_1^2-K_0^2)K_4+3K_2^2K_4=0,\label{eq3} \\
&&r^2 \pro_t^2 K_0-r^2 \pro_r^2 K_0-\pro_\theta^2 K_0+2r(\pro_t K_1-\pro_r  K_0)\nn \\
&&~~+\cot\theta (\pro_t K_2-\pro_\theta K_0)+ \dfrac{9}{2}\csc^2\theta K_4^2 K_0=0, \label{eq4}
\eea
 and 
one quadratic constraint 
\bea
&&2 r^2 (K_0\pro_t K_4-K_1 \pro_r K_4 )+ K_2 (\cot \theta K_4-2 \pro_\theta K_4)\nn \\
&&~~~~~+K_4 (-\pro_\theta K_2 +r^2 (\pro_t K_0-\pro_r K_1))=0.  \label{constr1}
\eea
A total Lagrangian ${\cal L}_{YM}+{\cal L}_{gf}$ takes a final form
\begin{align}
&{\cal L}_{\rm tot}={\cal L}_0(K)+{\cal L}_{\rm gf}, \nn \\
&{\cal L}_0(K)=\dfrac{1}{2r^2}\Big [r^2(\pro_t K_1 -\pro_r K_0)^2-(\pro_\theta K_1)^2+(\pro_\theta K_0)^2\Big ]\nn\\
&+\dfrac{1}{2 r^2} \Big [ \pro_t K_2 (\pro_t K_2- 2\pro_\theta K_0) - \pro_r K_2(\pro_r K_2-2\pro_\theta K_1) \Big ]\nn \\
&+\dfrac{3}{4r^4 \sin^2 \theta}\Big [ r^2 ((\pro_t K_4)^2-(\pro_r K_4)^2) -(\pro_\theta K_4)^2 \Big] \nn \\
&      -\dfrac{3}{4 r^4\sin^2 \theta} \Big [K_4^2 (K_2^2+r^2 (K_1^2-K_0^2)) \Big ]. \label{Lred}
\end{align}
The Lagrangian can be treated as a non-linear realization of $U(1)$ gauge theory for the vector field $K_{0,1,2,3}$
in the Lorenz type gauge, and it reproduces the equations (\ref{eq1}-\ref{eq4}).

It has been proved that stationary solutions satisfying equations (\ref{eq1}-\ref{eq4})
are stable under vacuum gluon fluctuations \cite{plb2018}.
The solutions are the only known classical solutions which  describe a stable vacuum in a pure Yang-Mills theory in quasiclassical 
approximation. Each solution to equations (\ref{eq1}-\ref{eq4}) implies a one parametric family of
solutions obtained by $SO(2)$ global rotation (\ref{residual}).
Since the global symmetry represents a subgroup of the initial color group $SU(2)$ we conclude that
$SU(2)$ Yang-Mills theory possesses a degenerated vacuum which leads to 
spontaneous color symmetry breaking and absence of color confinement phase.

\section{Weyl symmetric ansatz in $SU(3)$ Yang-Mills theory with Dirac fermions}

\subsection{ A pure $SU(3)$ Yang-Mills theory}

A minimal  Weyl symmetric ansatz for stable axially symmetric stationary solutions in $SU(3)$ Yang-Mills theory 
has been constructed in \cite{PRL1}. In this section we derive the most general minimal Weyl symmetric ansatz and 
show that there are two types of Weyl symmetric ansatz in 
a pure $SU(3)$ Yang-Mills theory. However, as we will see later, the consistence with the Weyl symmetric structure for 
fermions implies uniqueness of the ansatz. 
We define a general Weyl symmetric ansatz by introducing the following non-vanishing field components of the gauge potential
$A_\mu^a(r, \theta,\varphi,t)$
 $(a=1,2,...8)$ in three color subspaces corresponding to $I,U,V$-types subgroups $SU(2)$
($i=0,1,2$ denotes the space-time coordinates $(t,r,\theta) $ respectively)
\begin{align} 
I:~~A_i^2&=a_1 K_i,     & A_\varphi^1&=q_1 K_4,\nn \\
U:~A_i^5&=-Q_i,  & A_\varphi^4&=Q_4, \nn \\
V:~A_i^7&=S_i, & A_\varphi^6&=S_4,\nn \\
{\Ab}_\varphi^p&=A_\varphi^\alpha r_\alpha^{\,p}, & 
A_\varphi^3&=K_3, & A_\varphi^8&= K_8, \label{SU3DHN} 
\end{align}
where $a_1,q_1$ are number parameters, index $p=1,2,3$ denotes $I,U,V$ components of vectors 
in the Cartan plane $\{T^3, T^8\}$, and $r_\alpha^{\,p}$ $(\alpha=3,8)$ are components of 
symmetric $I,U,V$-root vectors 
$\bfr^{1}=(1,0)$, $\bfr^{2}=(-1/2,\sqrt 3/2)$, $\bfr^{3}=(-1/2,-\sqrt 3/2)$.
The ansatz is explicit  invariant under Weyl transformations which are presented by a symmetric permutation group $S_3$
acting on fields located in $I,U,V$ sectors.
One has three Weyl multiplets (numerated by index $i=(0,1,2)$) of off-diagonal fields, $\{K_i, Q_i,S_i\}$, 
each of them forms a reducible representation of the symmetric group $S_3$. 
Linear combinations of two Abelian
fields $A_\varphi^{3,8}$ corresponding to Cartan subalgebra generators  form a triplet of $I,U,V$-vectors,
$\{A_\varphi^I,A_\varphi^U,A_\varphi^V\}$,
which satisfy a constraint
\bea 
A_\varphi^I+A_\varphi^U+A_\varphi^V=0.
\eea
The equation is explicit  invariant under permutations of $I,U,V$-fields and defines a two-dimensional standard irreducible representation 
of $S_3$. 
The remaining fields $\{K_4, Q_4, S_4\}$  forms a three-dimensional  reducible representation of $S_3$.  A pair of Cartan generators
${\bf T}=(T^3,T^8)$, acts on generators from the coset $su(3)/u_3(1)\times u_8(1)$ in adjoint representation 
 in the Cartan basis as follows
 \bea 
&&{[\bf T}, T_+^p]= \bfr^p T_+^p. \label{eigenvalue1}
\eea
The eigenvalues $\bfr^p$ composed from the root vectors define color charges  $C^p$ of off-diagonal  $I,U,V$ fields
(we set $g=1$) 
\bea
 &&g^I=(1,0), \nn \\
 && g^U=(-\dfrac{1}{2},\dfrac{\sqrt 3}{2}),\nn \\
 &&g^V=(-\dfrac{1}{2},-\dfrac{\sqrt 3}{2}).\label{eigenvalue2}
 \eea
 $I,U,V$ components of the Abelian fields have the same charges. Total color charges of all Weyl multiplets
 vanish identically. 
One can check consistence of the ansatz  with all equations of motion corresponding to the $SU(3)$ Yang-Mills Lagrangian with
Lorenz type gauge fixing terms ${\cal L}_{gf}$ (\ref{gfterm}) added in each $I,U,V$ sector.

One can reduce the general ansatz to a minimal Weyl symmetric ansatz by extracting irreducible representations.
Due to Lorentz covariance the reduction of the general ansatz to a minimal one can be
done by imposing the following constraints
\bea
Q_i&=a_2 K_i,~~~~~&S_i=a_3 k_i \nn \\
Q_4&=q_2 K_4,~~~~~~ & S_4= q_3 K_4, \nn \\
K_3&=c_3  K_4, ~~~~~~~& K_3=c_8 K_8. \label{reduction1}
 \eea
Consistence of the full ansatz (\ref{SU3DHN}, \ref{reduction1})
 with all equations of motion is provided by the following values of number parameters $a_i,q_i,c_3,c_8$
\bea
&&a_i=b_i=1, \label{ab} \\
&&q_2=r^2_3+r^2_8=\Big (-\frac{1}{2}+\frac{\sqrt 3}{2}\Big ), \label{q4} \\
&&q_3=r^3_3+r^3_8= \Big (-\frac{1}{2}-\frac{\sqrt 3}{2}\Big ), \label{q6} \\
&&c_8=1, \label{c8} \label{c8} \\
&&c_3^I= -\dfrac{\sqrt 3}{2}, ~~~~~c_3^{II}=\sqrt 3, \label{c3} 
\eea
where the numbers $q_i$ are composed from the roots.
All parameter values are defined uniquely except the parameter $c_3$
which admits two different values, $c_3^{I,II}$.
In a pure Yang-Mills theory both values lead to the same quantum theory due to normalization freedom of 
 the fields $K_{\hat i}$ $(\hat i=0,1,2,4)$. With this, the $SU(3)$ Yang-Mills Lagrangian reduces 
 to a simple form in terms of four independent fields $K_{\hat i}$
\begin{align}
&{\cal L}_{red}=\dfrac{3}{2r^2}\Big [r^2(\pro_t K_1 -\pro_r K_0)^2-(\pro_\theta K_1)^2+(\pro_\theta K_0)^2\Big ]\nn\\
&+\dfrac{3}{2 r^2} \Big [ \pro_t K_2 (\pro_t K_2- 2\pro_\theta K_0) - \pro_r K_2(\pro_r K_2-2\pro_\theta K_1) \Big ]\nn \\
&+\dfrac{9}{4r^4 \sin^2 \theta}\Big [ r^2 ((\pro_t K_4)^2-(\pro_r K_4)^2) -(\pro_\theta K_4)^2 \Big] \nn \\
&      -g_4^2\dfrac{1}{r^4\sin^2 \theta} \Big [K_4^2 (K_2^2+r^2 (K_1^2-K_0^2)) \Big ]. \label{Lred}
\end{align}
The reduced Lagrangian does not contain cubic interaction terms which disappear due to mutual cancellation, one has only
 a quartic interaction
with a Weyl invariant coupling constant $g_4$
\bea
g_4^2=\dfrac{3}{2}(q_1^2+q_2^2+q_3^2-q_1 q_2 -q_2 q_3-q_3 q_1)=\dfrac{27}{4}.
\eea
The Lagrangian ${\cal L}_{red}$ is equivalent to $SU(2)$
Lagrangian ${\cal L}_0$, (\ref{Lred}), up to rescaling  $K_i\rightarrow 1/\sqrt 3 K_i$. So the minimal 
Weyl symmetric $SU(3)$ Yang-Mills theory
has the same equations and solutions as ones in $SU(2)$ gauge theory. The difference is that $SU(2)$ solutions 
admit the global $SO(2)$ color symmetry, whereas $SU(3)$ solutions are non-degenerate and form  invariant space 
under all Weyl color transformations.

Let us consider color properties of solutions defined by the ansatz (\ref{SU3DHN}, \ref{reduction1}).
The color structure of solutions is determined by  Weyl representations which are described  by representations 
of the symmetric permutation group $S_3$.
The same values of parameters $a_i,b_i$, (\ref{ab}) imply that $I,U,V$ components of the off-diagonal fields $\{K_i,Q_i,S_i\}$ 
are identical to each other
\bea
K_i=Q_i=S_i.
\eea

Due to this one has three one dimensional irreducible Weyl representations $\{(K_i,K_i,K_i)=(1,1,1) K_i \}$ 
corresponding to three irreducible singlet representations $\{\Gamma_1\}_i$
of the symmetric group $S_3$. Therefore we have three Weyl symmetric color singlet fields $K_i$.
Parameters $q_{1,2}$ in (\ref{q4},\ref{q6}) are made from the root vectors $\bfr^2, \bfr^3$ implying an equation
\bea 
K_4+Q_4+S_4=0.
\eea
This equation defines a two dimensional plane in the space spanned by $I,U,V$ components $(K_4,Q_4,S_4)$.
The equation is explicit invariant under permutations and defines a standard two-dimensional irreducible representation
 $\Gamma_2(K_4)$ of $S_3$ which in general contains two independent fields. In our case this irreducible representation is realized 
in terms of only one independent field $K_4$, i.e., it is a singlet. In a similar manner the $I,U,V$ combinations of the 
Abelian fields $A_\varphi^{3,8}$ form an equivalent  standard two-dimensional irreducible representation $\Gamma_2(K_3)$. 
Consistence of the ansatz with equations of motion implies that
 there is only one independent Abelian field in the Cartan plane,$A_\varphi^3=A_\varphi^8\equiv K_3$,
  (\ref{reduction1}, \ref{c8}).
Finally, two representations $\Gamma_2(K_4)$ and $\Gamma_2(K_3)$ are equivalent due to the constraint 
 (\ref{reduction1}, \ref{c8}).
Let us show that field $K_4$ (and $K_3$ equivalently)  represents a Weyl invariant field.
We consider eigenvalues of a Lie algebra valued Abelian vector field,
 $\bfA_\mu = (A_\mu^3 T^3,~A_\mu^8 T^8)$, acting in adjoint representation 
 in the Cartan basis. In a case of equalled Abelian fields, $A_\varphi^3=A_\varphi^8$, 
 one can find
 \bea 
&&[\bfA_\varphi^p, T_+^p]=K_3 \bfr^p T_+^p.
\eea
The eigenvalues $K_3 \bfr^p$ match the symmetric roots $\bfr^p$ with a normalization function $K_3$, so that 
the $I,U,V$-components of the Abelian field  possess the same Weyl symmetry as the roots, and the field $K_3$ 
represents a Weyl invariant normalization function which produces singlet quantum states for Abelian gluon
(more details on singlet structure of representations of the permutation group are presented in the Appendix). 

We conclude, the Weyl symmetry implies that one has three color singlet fields $K_i$ for off diagonal fields, corresponding to 
singlet irreducible representations $\{\Gamma_1\}_i$ and one color singlet Abelian field $K_3$ (or $K_4$ equivalently) 
entering two equivalent standard irreducible representations $\Gamma_2(K_4)$ and $\Gamma_2(K_3)$.

It is remarkable, {\it the Weyl symmetric ansatz (\ref{SU3DHN},\ref{reduction1}) leads to reduction of the initial eight 
dimensional $SU(3)$ octet $A_\mu^a$ to four irreducible  singlet Weyl representations for four fields $K_i, K_4$ belonging
to invariant non-intersecting subspaces under color Weyl transformations (permutations of $S_3$)}. 

So far we have considered only transformation properties of field configurations satisfying the Weyl symmetric ansatz 
(\ref{SU3DHN}, \ref{reduction1}) under Weyl transformations (or permutations of the symmetric group $S_3$).
We did not take into account dynamical properties of solutions which determine the number of dynamical degrees of freedom.
These issues are considered in \cite{PRL1, PRD1} where it has been demonstrated that among four fields $K_{\hat i}$
there are only two dynamical propagating modes, $K_2$ and $K_4$. This can be easily understood from the fact
 that a final reduced Lagrangian is identical to $SU(2)$ Lagrangian
${\cal L}_{tot}$, (\ref{Lred}), which represents a non-linear $U(1)$ gauge theory. Due to the presence of a 
local $U(1)$ symmetry one dynamical degree of freedom 
represents a pure gauge field, and another degree of freedom is removed on-shell, leaving only two dynamical 
degrees of freedom, $K_{2,4}$. Moreover, one would expect that magnetic type non-Abelian solution should contain 
only one 
dynamical degree of freedom, as in the Maxwell theory the vector spherical harmonics of magnetic and electric types
 correspond to
 two possible photon polarizations, one polarization mode for magnetic and one mode for electric type harmonic.
 Indeed, the magnetic type non-Abelian Weyl symmetric solution admits only one independent dynamic field which can be chosen
 as Abelian field $K_3$. A source of such a  phenomenon  is very non-trivial and related to another important requirement for 
 physical solutions - solutions must admit localization of corresponding quantum states in a finite space regions. 
 Indeed, it has been proved \cite{PRD1}, that solutions for a given set of quantum numbers admit energy conservation 
 law inside a finite  space region if amplitudes of fields $K_{2,4}$ are correlated, as a consequence, one has finally 
 only one independent field $K_3$ \cite{PRD1}. In a similar manner one can consider electric type solutions due to 
 electric-magnetic duality \cite{PRD1}.
 
\subsection{Weyl symmetric structure of $SU(3)$ Yang-Mills theory with fermions}

Let us consider Weyl symmetric structure of  a coupled system of 
gluon and quarks presented by one $SU(3)$ complex triplet of Dirac spinor fields  (quarks) in fundamental representation
(\ref{eqA}, \ref{eqPsi}).

A commonly accepted simple Abelian projection with  two independent Abelian fields $A_\mu^{3,8}$ corresponding to 
the Cartan generators immediately results in three independent Dirac equations for three independent color quarks
\bea
   \Big[ i \gamma^\mu \pro_\mu -\dfrac{i g}{2}  \sum_p A_\mu^\alpha w_\alpha^p-m\Big]\Psi_i=0,\label{usualqeq}
\eea
where color charges of quarks are defined by components $w^p_\alpha$ ($\alpha=3,8$) of weight vectors 
$\bfw^p=\{(1, 1/\sqrt 3), (-1,1/\sqrt 3), (0,-2/\sqrt 3)\}$.
It is important to observe, that a Weyl symmetric ansatz for the whole set of components of the gauge potential
leads to the Weyl symmetric ansatz (\ref{SU3DHN}, \ref{reduction1}) which contains only one Abelian field.
To preserve a Weyl symmetry of the total Lagrangian including fermions, we must apply a Weyl symmetric
ansatz  which is consistent with a full non-linear structure of the Yang-Mills Lagrangian.
Substituting the Weyl symmetric ansatz (\ref{SU3DHN}, \ref{reduction1}) into the equation for quark,(\ref{eqPsi}), 
one obtains a drastically different equation
\bea
 \Big[ i \gamma^\mu \pro_\mu-m+\dfrac{g}{2} \gamma^\mu  A_\mu G 
                +\dfrac{g}{2}\gamma^\mu K_\mu Q\Big]\Psi=0, \label{GQeqns} 
\eea
where $A_\mu=\delta_{\mu 3}K_3$,  $K_\mu=\delta_{\mu i} K_{i}$($i=0,1,2$).
A color charge matrix $Q$ defines
interaction of quarks  with off-diagonal gluon fields $K_i$ 
\bea
 Q=\begin{pmatrix}
   0&-i a_1& i a_2\\
  i a_1&0&-i a_3\\
   -i a_2 &i a_3&0\\
 \end{pmatrix}, 
 \eea
 and a color charge matrix $G$ determines interaction  with the Abelian gluon field $K_3$
\bea
  G=\begin{pmatrix}
   \tw^1&~~ c_3^{-1}( r^3_3+r^3_8) &~~c_3^{-1}( r^2_3+r^2_8)\\
   c_3^{-1}( r^3_3+r^3_8)&~~\tw^2& ~~c_3^{-1}( r^1_3+r^1_8)\\
  c_3^{-1}( r^2_3+r^2_8) &~~c_3^{-1}( r^1_3+r^1_8)&\tw^3\\
 \end{pmatrix}.\nn \\
 \label{Gmatrix}
 \eea
Substituting the value $c_3=c_3^I$ and explicit expressions for the root vectors $r^p_\alpha$
the charge matrix $G$ obtains a simple form which is explicit invariant under permutations of matrix elements in each row and column
\bea
  G=\begin{pmatrix}
   \tw^1&\tw^3&\tw^2\\
  \tw^3&\tw^2&\tw^1\\
   \tw^2&\tw^1&\tw^3\\
 \end{pmatrix}.
\eea
This provides a Weyl symmetric structure of the equation for quarks, and implies irreducible Weyl representations
which describe singlet quarks.
A total sum of elements in each row and column vanishes identically due to properties of weights.
This defines a one dimensional irreducible representation of  the permutation group $S_3$ with a basis color eigenvector 
$u^0=(1,1,1)$
\bea
\Psi_0=u^0 \psi_0, \label{qneutral}
\eea
where $\psi_0$ is a Dirac spinor field  with space-time coordinate dependence.
The eigenvector $\Psi_0$ corresponds to a zero eigenvalue, i.e., zero color charge.
One can verify that the charge matrix $Q$ has the same zero mode $\Psi_0$, so the neutral quark does not interact with 
 Abelian and off-diagonal gluons. 
The neutral quark satisfies a free Dirac equation and determines a color singlet vacuum structure. In a particular 
it describes free vacuum quark condensate.
One can verify that equation for gluon, (\ref{eqA}), decouples from equations for quarks, (\ref{eqPsi}),
 so the coupled system of non-linear equatons
for gluon  and quarks admits solutions for non-interacting free gluon and quark condensates. The color charge matrix $G$ 
has other two non-zero eigenvalues 
\bea 
\lambda_\pm=\pm \tg=\pm\sqrt 6,
\eea
where
\bea
\tg^2=(\tw^1)^2+(\tw^2)^2+(\tw^3)^2-\tw^1\tw^2-\tw^2\tw^3-\tw^3 \tw^1  \nn\\
\eea
is a Weyl invariant color charge. One can find corresponding two color eigenvectors
\bea
u^\pm=\begin{pmatrix}
   \tw^1\tw^3+\tw^2(\pm\tg-\tw^2)\\
 \tw^2\tw^3+\tw^1(\pm\tg-\tw^1))\\
   \tw^1\tw^2+\tw^3(\pm\tg-\tw^3)+\tg^2 \\
 \end{pmatrix},\label{upm}
 \eea
which form an orthogonal basis for a two dimensional standard irreducible representation $\Gamma_2$.
Both eigenvectors satisfy an equation for the same two-dimensional plane $P^2$
\bea 
u_1^\pm+u_2^\pm+u_3^\pm=0.
\eea
Note that each color eigenvector $u^\pm$ does not form an irreducible one dimensional representation of $S_3$
(see Appendix for more details).

It is important to stress, that 
each quark triplet $\Psi^\pm=u^\pm \psi^\pm(\vec r,t)$ contains a color vector $u^\pm$ which belongs
to the irreducible two-dimensional representation $\Gamma_2$. However,  the non-zero mode
$\Psi^\pm$ contains  only one field variable $\psi^\pm$, so it is a singlet field. 
One can verify that two sets of vectors obtained by 
permutations of components of two basis vectors $u^\pm $ form two
Weyl invariant sets which do not intersect each other, 
even though they form a complete basis in the two dimensional plane $P^2$. 
Therefore, two quark triplets $\Psi^\pm=u^\pm \psi^\pm(\vec r,t)$ represent two independent  irreducible singlet 
representations with respect to a number of independent field variables, two single quarks,
which interact to Abelian gluon with a Weyl invariant  color charge $\pm \tg$.  Interaction of non-zero quark modes 
with off-diagonal gluons is described by the charge matrix $Q$
which has a different non-zero Weyl invariant charge
\bea
\hg^2=a_1^2+a_2^2+a_3^2=3.
\eea

The second type of Weyl symmetric solutions defined by the parameter value $c_3^{II}=\sqrt 3$ does not
provide Weyl symmetric form for the color charge matrix $G$. So that one has a unique Weyl symmetric  structure with
a unique non-degenerate vacuum. We conclude,  {\it the structure of color confinement and quantum states 
in  $SU(3)$ Yang-Mills theory is determined completely by the Weyl symmetry which 
is the only available color symmetry resulting in color singlet representations for gluons and quarks}. 
Definitions for color gluons and quarks defined as members of $SU(3)$ color multiplets represent artifacts of the applied 
perturbation theory and simple Abelian projection which are not consistent 
with the whole non-Abelian structure of the total Lagrangian including gluons and quarks.

\section{Weyl symmetric structure of $SU(4)$ Yang-Mills theory}

We use a standard Weyl basis $\{\lambda^i\}$ $(i=1,2,...15)$ in the Lie algebra $su(4)$. 
 The Weyl group of $SU(4)$ is presented by a symmetric permutation
group $S_4$. We consider a minimal Weyl symmetric ansatz for time dependent axially symmetric fields 
of magnetic type.

A Cartan subalgebra of group $SU(3)$ contains three generators $T^{3,8,15}$, so that for magnetic type fields
one has corresponding three independent Abelian magnetic potentials $A_\varphi^{3,8,15}$.
One has six root vectors $r^p_\alpha$ $(p=1,2...6)$, $(\alpha=3,8,15)$, defined as eigenvalues of Cartan generators acting on complex generators
of the coset space $SU(4)/U_3(1)\times U_8(1)\times U_{15}(1)$ in adjoint representation
\bea
&&\bfr^1=(1,0,0),\nn \\
&&\bfr^2=(-\dfrac{1}{2} ,\dfrac{\sqrt 3}{2},0),\nn \\
&&\bfr^3=(-\dfrac{1}{2},-\dfrac{\sqrt 3}{2}, 0), \nn \\
&&\bfr^4=(\dfrac{1}{2}, \dfrac{1}{2 \sqrt 3}, \dfrac{2}{\sqrt 6}), \nn \\
&&\bfr^5=(-\dfrac{1}{2}, \dfrac{1}{2 \sqrt 3}, \dfrac{2}{\sqrt 6}), \nn \\
&& \bfr^6=(0,\dfrac{1}{\sqrt 3}, -\dfrac{2}{\sqrt 6}).
\eea
A general Weyl symmetric ansatz contains three
Abelian magnetic fields $A_\varphi^{3,8,15}$ in the Cartan space,
additional six independent Abelian fields $A_\varphi^{1,4,6,9,11,13}$ in the coset subspace 
spanned by generators $T^{1,4,6,9,11,13}$ and
eighteen off-diagonal fields  $K_i, Q_i, S_i, P_i, T_i, R_i$ ($i=0,1,2$)
\bea
&A_i^2=p_1 K_i,~~&A_i^5=Q_i,~~A_i^7=S_i,  \nn \\
&A_i^{10}=P_i,~~&A_i^{12}=T_i,~~A_i^{14}=R_i,  \nn \\
&A_\varphi^1=q_1 K_4,~~&A_\varphi^4=Q_4,~~A_\varphi^6=S_4,  \nn \\
&A_i^9=P_4,~~&A_i^{11}=T_4,~~A_i^{13}=R_4,\nn \\
&A_\varphi^3=K_3,~~&A_\varphi^8=K_8,~~A_\varphi^{15}=K_{15},  \label{SU4DHN} 
\eea
where $p_1,q_1$ are number parameters. One can verify that ansatz (\ref{SU4DHN})
is consistent with Yang-Mills equations of motion, and symmetric under Weyl permutations
in six dimensional spaces spanned by $T^{1,4,6,9,11,13}$ and $T^{2,5,7,10,12,14}$
and in the root space spanned by six linear combinations of Cartan generators $r^p_\alpha T^\alpha$.

One can reduce the ansatz and construct a minimal 
Weyl symmetric ansatz with four independent fields $K_t=K_0, K_r=K_1, K_\theta=K_2, K_\varphi=K_4$
by imposing additional reduction constraints
\bea
&K_3=c_3 K_4, ~~&K_8=c_8 K_4, ~~K_{15}=c_{15} K_4, \nn \\
&Q_i=p_2 K_i,~~&S_i=p_3 K_i,\nn \\
&P_i=p_4 K_i,~~&T_i=p_5 K_i,~~R_i=p_6 K_i,\nn \\
&Q_4=q_4 K_4,~~&S_4=q_6 K_4,\nn \\
&P_4=q_9 K_4,~~&T_4=q_{11} K_4,~~R_4=q_{13} K_4, \label{SU4reduction}
\eea
where $c_{3,8,15},p_{2-6},q_{2-6}$ are number parameters.
In a case of $SU(3)$ Yang-Mills theory the parameters $q_{1,4,6}$
are defined  by root vectors which provide a proper definition of 
an irreducible Weyl group representations. In a case of $SU(4)$ Yang-Mills theory
the parameters $q_\alpha$ can not be given by roots of $SU(4)$ Lie algebra
since a total sum of roots does not vanish contrary to the case of $SU(3)$ roots, 
and, as a consequence, one meets
an obstacle with constructing a standard five dimensional irreducible Weyl representation
by setting a constraint
\bea
K_4+Q_4+S_4+P_4+T_4+R_4=0.
\eea
To resolve this problem we treat the parameters $q_\alpha, p_\alpha$ as free parameters providing 
a Weyl symmetric structure of the charge matrices $G,Q$ in the Dirac equation for quarks in fundamental representaion
of  $SU(4)$ with applied Abelian projection defined by 
the ansatz (\ref{SU4reduction})
\bea
 \Big[ i \gamma^\mu \pro_\mu-m+\dfrac{g}{2} \gamma^\mu  A_\mu G +\dfrac{g}{2} \gamma^\mu K_\mu Q
               \Big]\Psi=0, \label{GQeqns} 
\eea
where $A_\mu=\delta_{\mu 3}K_4$, $K_\mu=\delta_{\mu i} K_i$,   the color charge matrix $G $ reads
\bea
  G=\begin{pmatrix}
   \tw^1&q_1&q_4&q_9\\
  q_1&\tw^2&q_6&q_{11}\\
  q_4&q_6&\tw^3&q_{13}\\
  q_9&q_{11}&q_{13}&\tw^4\\
 \end{pmatrix},\label{G}
\eea
with diagonal elements $\tw^\alpha$ representing linear combinations of weights of $su(4)$ Lie algebra
\bea 
\tw^1=c_3\cdot 1+&c_8\dfrac{1}{\sqrt 3}+&c_{15} \dfrac{1}{\sqrt 6}, \nn \\
\tw^2=-c_3\cdot 1+&c_8\dfrac{1}{\sqrt 3}+&c_{15} \dfrac{1}{\sqrt 6}, \nn \\
\tw^3=  ~~~~~~~~~~     &c_8\dfrac{-2}{\sqrt 3}+&c_{15} \dfrac{1}{\sqrt 6}, \nn \\
\tw^4=~~~~~~~~~&~~~~~~~~&c_{15} \dfrac{-3}{\sqrt 6}.
\eea
A color charge matrix $Q$ providing coupling constants of off-diagonal gluons with quarks
has the following form
\bea
  Q=\begin{pmatrix}
   0  & -i p_1 & -i p_2 &-i p_4  \\
  i p_1 &  0   &-i p_3 &-ip_5\\
  i p_2 & i p_3 &  0   &-ip_6\\
  ip_4 & ip_5 & ip_6  &   0  \\
 \end{pmatrix},\label{Q}
\eea
One can set  $q_1=1, p_1=1$ due to normalization freedom of the fields $K_i, K_4$.
Weyl symmetric structure of the Dirac equation implies that a total sum of matrix elements
in each row and column in matrices (\ref{G},\ref{Q}) must vanish.
Consistence equations for the matrix $Q$ can be easily found
\bea
&&p_3=-1-p_2,\nn \\
&&p_4=-1-p_2,\nn \\
&&p_5=2+p_2,\nn \\
&& p_6=-1,\nn 
\eea
Corresponding consistence equations for matrix $G$ allow to find expressions for parameters $c_3, c_8, c{15}, q_{13}$
 in terms of remaining parameters. With this  one can solve analytically all constraints arising from consistence of the ansatz  with 
 all equations of motion.
Final result is the following
\bea
&&c_3=\dfrac{p_2(2+p_2)(-1+p_2-q_9)-2(1+q9)}{2 p_2(1+p_2)},\nn \\
&&c_8=\dfrac{2(1+q_9)-p_2(2(1+q_9)+p_2(5+p_2+3q_9))}{2 \sqrt 3 p_2(1+p_2)},\nn \\
&&c_{15}=\dfrac{(2+p_2)(2(1+q_9)+p_2(3+p_2+3q_9))}{\sqrt   6 p_2 (1+p_2)},\nn 
\eea
\bea
&&q_4=-\dfrac{2+p_2)(1+p_2+q_9)}{2 (1+p_2)},\nn \\
&&q_6=-\dfrac{1}{p_2}(2+2p_2+2q_9+p_2q_9),\nn \\
&&q_{11}=\dfrac{(2+p_2)(1+p_2+q_9)}{2(1+p_2)}\nn \\
&&q_{13}=\dfrac{1}{p_2}(2+p_2+2q_9),
\eea

The solution forms a two-parametric family 
defined by two arbitrary numbers $p_2,q_9$. 
The parameters satisfy two relationships
\bea
&&q_1+q_4+q_6+q_9+q_{11}+q_{13}=0, \\
&& p_1+p_2+p_3+p_4+p_5+p_6=0,
\eea
which provide exisence of standard irreducible representations $\Gamma_5$.
One can find that color charge matrix $G$ has two zero modes, one of which belongs to singlet irreducible representation
containing a free  zero quark mode. Other two color quark modes have opposite color charges.
The matrix $Q$ has also two zero modes for a free color singlet quarks, one of which coincides with the zero mode of the matrix $G$.
Other two non-zero quark modes have opposite color charges and can be decomposed in the basis of non-zero modes of the
charge matrix $G$.
The obtained ansatz provides a Weyl symmetric structure of the gluon fields $K_i, K_4$ and for quarks. However, 
due to presence of two free parameters $p_2,q_9$ the ansatz defines solutions which are degenerated.
In a general, this implies a spontaneous color symmetry breaking which excludes color  confinement phase.

\section{Discussions}

A similar analysis can be performed for $SU(5)$ Yang-Mills theory.
We construct first a Weyl symmetric ansatz in a pure Yang-Mills theory.
A general ansatz for magnetic type time-dependent solutions contains
four Abelian potentials $A_\varphi^{3,8,15,24}$ corresponding to the maximal Abelian subgroup
of $SU(5)$, ten Abelian potentials $A_\varphi^p$ with an internal index 
$p=(1,4,6,9,11,13,16,18,20,22)$ which corresponds to coset generators $T^{1,4,6,9,11,13,16,18,20,22}$
and it is used also for counting the root vectors,
and ten off-diagonal vector fields $A_i^{\dot p}$ ($i=0,1,2$) 
corresponding to index values $\dot p=(2,5,7,10,12,14,17,19,21,23)$ corresponding to
coset generators $T^{2,5,7,10,12,14,17,19,21,23}$.
A minimal ansatz is reduced by imposing reduction constraints
\bea
&&K_\alpha=c_\alpha K_4, \nn \\
&&A_i^{\dot r}=p^{\dot r} K_i, \nn \\
&& A_\varphi^r=q^r K_4,, \label{SU4reduction}
\eea
where index $\alpha=3,8,15,24$ denotes for generators of the Cartan subalgebra,
and $K_4$ is the only Abelian  magnetic potential.
Consistence of the minimal ansatz with all equations of motion 
leads to constraints on parameters $c_\alpha, p^{\dot r}, q^r$.
Additional constraints appear from the requirement that equation for 
a quark must possess Weyl symmetric structure. 

The number of independent constraints equals the number of free parameters, so we resolve all
constraints numerically and have found a following solution for the parameters
\begin{align}
&p_1=-1,~~~~~~&q_1=1,\nn \\
&p_2=-0.138325,    &q_2=0.011256,\nn \\
&p_3=-0.535107,    &q_3=0.082786,\nn \\
&p_4=-0.277493,     &q_4=0.011342,\nn \\
&p_5=-0.247661,     &q_5=0.154373,\nn \\
&p_6=0.182746,      &q_6=0.199056,\nn \\
&p_7=0.535065,      &q_7=-0.052219\nn \\
&p_8=0.577338,      &q_8=-0.285323,\nn \\
&p_9=-0.366178,     &q_9=-0.406182,\nn \\
&p_{10}=0.027693,    &q_{10}=-0.715090, \nn 
\end{align}
\bea
&&c_3=-0.009271, \nn \\
&&c_8=-0.620184, \nn \\
&&c_{15}=-0.583812,\nn \\
&&c_{24}=-1.153294, \nn 
\eea
The color charge matrix $G$ has three zero modes, and two non-zero modes
with effective coupling constants $\lambda=\pm 1.412....$.
The obtained parameter values provides a Weyl symmetric 
structure of the minimal ansatz for a pure Yang-Mills theory. In a case of presence of fermions the number of 
consistence equations providing Weyl symmetric structure becomes more than number of free parameters.
We have not found Weyl symmetric ansatz in a case of  including fermions, however, we  can not exclude completely its existence 
since the consistence equations have a high non-linear structure and there might be some special solutions for the parameters
which are missed during numeric solving.

In conclusion, we have considered an ansatz for Weyl symmetric classical stationary solutions in $SU(N)$ ($N=2,3,4,5$) Yang-Mills theory
which describe color singlet one particle states for gluons and quarks.
Previous approaches to construction of a gauge
    invariant vacuum or gauge invariant quantum states, fail because they were based on a seemed natural assumption 
    that color symmetry of quantum states, gluons and quarks, is provided by a global color group $SU(3)$.
We have demonstrated that a Weyl symmetric ansatz  in $SU(2)$ and $SU(4)$ Yang-Mills theory admits a global color symmetry 
which leads to degeneracy of solutions, as a consequence, the vacuum in these theories is degenerate.
Selection of a single vacuum in general leads to spontaneous color symmetry breaking 
which excludes color confinement phase in the theory. In a case of $SU(5)$ Yang-Mills theory we did not find 
an ansatz with a minimal set of four gluon fields $K_i, K_4$
which provides Weyl symmetry of the total Lagrangian including fermions.
We suppose that Weyl symmetric color singlet solutions for gluons and quarks exist only in $SU(3)$ Yang-Mills theory, i.e., in QCD. 
Our results reveal a non-trivial color structure of solutions necessary for construction of color singlet one-particle
quantum states for gluons and quarks. Abelian solutions 
provide color invariant structure of the vacuum, which determines the color confinement phase in QCD \cite{polyakov77, thooft81}.
  This opens a novel way to description of hadron structure.

\acknowledgments

This  work is supported by Chinese Academy of Sciences  (PIFI Grant No. 2019VMA0035),
National Natural Science Foundation of China (Grant No. 11575254),
and by Japan Society for Promotion of Science (Grant No. L19559).

\newpage
{\bf Appendix} \\

We use main definitions and results from the group theory related to
representations of the symmetric group $S_N$.

A representation of the symmetric group $S_N$ is defined as a group homomorphism of the group $S_N$
to the group of automorphisms of a vector space $V$. In other words the symmetric group $S_N$ can be realized by
a symmetric group $S_N$ acting by permutations on vector components of a vector $\vec V=(V_1, V_2,...V_N)$.
A one-dimensional irreducible representation (the sign representation) $\Gamma_1$ is defined by a constraint imposed
on the vector components 
\bea
V_1=V_2=...=V_N\equiv V.
\eea
The vector $\vec V=(1,1,...,1)V$ is invariant (symmetric) under permutations of its components
and represents a singlet representation. 
A standard $N-1$-dimensional irreducible representation $\Gamma_{N-1}$ is defined by equation
\bea
V_1+V_2+....+V_N=0.
\eea
The equation defines a $N-1$ dimensonal  hyperplane in $R^N$ which represents an invariant subspace under permutations 
of vector components. 
A dimension of the standard representation $\Gamma_{N-1}$ is defined by dimension of the hyperplane and equals $N-1$.
The standard representation is irreducible.

Let us compare representations of the unitary group $SU(3)$ which is a color group for QCD, and representations of the corresponding 
Weyl group which is a finite color subgroup of $SU(3)$.
A quark multiplet in fundamental representation of $SU(3)$ is given by three complex Dirac spinor functions
forming a fundamental color triplet
\bea
\Psi=(\psi_1, \psi_2, \psi_3), \label{qtriplet}
\eea
where $\psi_i$ are Dirac spinor functions depending on space-time coordinates.
The representation is irreducible and three-dimensional in color space. In addition one has three independent coordinate spinor functions which 
describe three independent particles, color quarks. 
In quantum Yang-Mills theory, after removing pure gauge fields by gauge fixing procedure, a color symmetry is given by a finite
color Weyl group, which is equivalent to permutation group $S_3$, which has different representations.
For instance, a complex quark triplet (\ref{qtriplet}) corresponding to the fundamental representation of $SU(3)$ color group
realizes a reducible representation of $S_3$.
It contains one dimensional irreducible representation $\Gamma_1$ 
with a basis vector
\bea 
\Psi_0=(1,1,1) \psi^0\equiv u^0 \psi^0
\eea
with a coordinate spinor function $\psi^0$. It is clear that representation describe one particle, i.e., it is a singlet representation
containing a neutral quark with zero color charge, (\ref{qneutral}).
Another irreducible representation is given by a standard two-dimensional representation defined by equation
\bea
\psi_1+\psi_2+\psi_3=0.
\eea
One has two independent spinor functions which describe two quarks, so the number of independent quarks is the same.
Despite on a mathematical fact that representation $\Gamma_2$ is irreducible and can not be decomposed further into irreducible
representations of smaller dimensions, we have shown in the paper that these two quarks correspond to two 
eigenvectors $u^\pm$, (\ref{upm}), of the color charge matrix $G$ with Weyl invariant color charges $\tg=\pm \sqrt 6$,
and the quarks are described by two complex triplets
\bea 
&&\Psi^+=u^+ \psi^+, \nn \\
&&\Psi^-=u^- \psi^-,
\eea
where color and coordinate parts are factorized.The color vectors $u^{0\pm}$ form a complete orthogonal basis vectors in the color space.
The color vectors $u^\pm$ are not singlets. Under permutations of their components each eigenvector generates a 
sextet consisting of six elements which lie in the two-dimensional plane and do not form a one-dimensional space. 
Elements of two sextets are  pairwise orthogonal and form two Weyl invariant non-intersecting sets.
Each quark triplet $\Psi^\pm$ contains one independent Dirac spinor  $\psi^\pm$
which describes a single quark.
Thus, each quark triplet $\Psi^\pm$ describes
one color singlet quark $\psi^\pm$. Note that the presence of a sextet of color vectors $\{u^+\}$ 
obtained by permutations of vector components 
$u_i^+$ does not imply that quark solution is degenerated since one has only one single quark described by 
Weyl invariant spinor $\psi^+$.

\end{document}